\documentclass[preprint,preprintnumbers,amsmath,amssymb]{revtex4}
\usepackage{graphicx}
\usepackage{graphics}
\usepackage{dcolumn}
\usepackage{bm}
\usepackage{amsfonts}
\usepackage{indentfirst}
\usepackage{setspace,lscape}
\usepackage{longtable}

 \setlongtables

\begin{document}

%\preprint{APS/123-QED}

\title{Lower Bound for LMC complexity measure}

\author{\'{A}. Nagy}\affiliation{Department of Theoretical Physics, University of
Debrecen,\\ H--4010 Debrecen, Hungary}

\author{K.~D.~Sen}\email{sensc@uohyd.ernet.in }
\affiliation{School of Chemistry, University of Hyderabad,
Hyderabad 500046, India}

\author{H.E. Montgomery Jr.}
\affiliation{Centre College, 600 West Walnut Street,Danville, KY
40422-1394, USA}

\date{\today}

\begin{abstract}
Lower bound for the shape complexity measure of
L\'opez-Ruiz-Mancini-Calbet (LMC), $C_{LMC}$, is derived.
 Analytical relations for simple examples of the harmonic oscillator, the
hydrogen atom and  two-electron 'entangled artificial' atom
proposed by Moshinsky are derived. Several numerical examples of
the spherically confined model systems are presented as the test
cases. For the homogeneous potential, $C_{LMC}$ is found to be
independent of the parameters in the potential which is not the
case for the non-homogeneous potentials.

\end{abstract}

\maketitle

%\noindent {\bf PACS:}31.10.+z;31.15.-p;31.30.Jv;31.30.-i .

\vfill\eject

\section{Introduction}

There are several statistical measures of complexity
\cite{lopez1995,land1998}.  A given measure becomes significant
when a rigorous bound on it is known to exist. In this letter, we
focus on the LMC(L\'opez-Ruiz-Mancini-Calbet) complexity
\cite{lopez1995}, $C_{LMC}$ , with the aim of deriving a general
lower bound for it. The lower bound is tested by presenting (a)
the analytical expressions for some simple systems: the harmonic
oscillator, the hydrogen atom and two-electron  atom proposed by
Moshinsky \cite{mos} and (b) the numerical calculations on the
spherically confined model one and two electron systems
\cite{Michels1,Sen1,Sen2,Sen3}.

Consider a $D$-dimensional distribution function $f({\bf r})$,
with $f({\bf r})$ nonnegative and {$\int f({\bf r}) d{\bf r}=1$;
$\bf r$ stands for $r_1, ..., r_D$. The Shannon entropy
\cite{shannon} and the Shannon entropy power are defined as
\begin{equation}
\label{c7} S_f = -\int f({\bf r})\ln f({\bf r}) d{\bf r} .
\end{equation}
\begin{equation}
\label{c8} H_f = e^{S_f} ,
\end{equation}
respectively. The so called disequilibrium $D$ has the form
\begin{equation}
\label{c9} D_f = \int f^2({\bf r}) d{\bf r} .
\end{equation}
The definition of the LMC complexity measure is \cite{lopez2002}
\begin{equation}
\label{c10}
 C_{LMC}= H.D
\end{equation}
  It is known \cite{lopez2002}  that the complexity
corresponding to probability distributions given by rectangular,
triangular, Gaussian and exponential functions in one-dimensional
position space is given by 1,~(2/3)$(e^{1/2})$,~$(e^{1/2})$/2, and
e/2, respectively. The rectangular probability distribution, by
definition, corresponds to the minimum statistical complexity. We
shall now derive the lower bound for $C_{LMC}$ corresponding to a
given one-electron density.

\section{Inequality for the LMC complexity}

To derive a lower bound for the LMC complexity we cite Theorem 2
of the paper of Y\'a\~nez,  Angulo and Dehesa \cite{yad1995}. The
position-space entropy ${\tilde S}_{\varrho}$ of an $N$-electron
system in a physical state characterized by the (normalized to
$N$) one-electron density $\varrho({\bf r})$ fulfills the
inequality
\begin{equation}
\label{c11} {\tilde S}_{\varrho} + <\ln{g({\bf r})}>  \le N ln{
\left ( \frac{\int g({\bf r})d{\bf r}}{N}  \right)} ,
\end{equation}
where $g({\bf r})$ is an arbitrary positive function. From the
relationship between the Shannon entropies coming from densities
normalized to 1 and N \cite{yad1995}:
\begin{equation}
\label{c12}
 S_f = \frac{{\tilde S}_{\varrho}}{N} + \ln{N}
\end{equation}
and taking $g=f^2$, we obtain the inequality
\begin{equation}
\label{c13}
 S_f +  \ln{\left (\int f^2({\bf r})d{\bf r} \right)} \ge 0
\end{equation}
From the definition of the LMC complexity (\ref{c10}) we obtain
the upper bound
\begin{equation}
\label{c14a}
 \ln{C_{LMC}} \ge 0
\end{equation}
or
\begin{equation}
\label{c14b}
 C_{LMC} \ge 1  .
\end{equation}
In the next section analytical expressions are presented for some
simple systems, such as the harmonic oscillator, the hydrogen atom
and  two-electron  atom proposed by Moshinsky.

\section{ Analytical Examples}

Consider first the one dimensional box problem i.e. the infinite
square well in one dimension: $V(x)=0$, if $|x|<a$ and
$V(x)=\infty$, if $|x|>a$. Even solutions are
\begin{equation}
\label{c39} \psi_{e,n} =  A \cos{\left (\frac{n \pi x}{2
a}\right)}  ,
\end{equation}
if $|x|<a$. Odd solutions are
\begin{equation}
\label{c39b} \psi_{o,n} =  B \sin{\left (\frac{n \pi x}{2
a}\right)}  ,
\end{equation}
if $|x|<a$. $n=1,2,...$, $A=B=1/a^{1/2}$ is the normalization
constant. The disequilibrium takes the form:
\begin{equation}
\label{c40} D =  \frac{3}{4a}
\end{equation}
for all eigenfunctions. The Shannon entropy can be written as
\begin{equation}
\label{c42} S_e = \ln{a} -1 - \int_{-1}^{1} \ln{\left (\cos{\left
(\frac{n\pi u}{2}\right )}du \right )} ,
\end{equation}
for the even solutions and
\begin{equation}
\label{c42b} S_o = \ln{a} -1 - \int_{-1}^{1} \ln{\left (\sin{\left
(\frac{n\pi u}{2}\right )}du \right )} ,
\end{equation}
for the odd solutions. Eqs. (\ref{c40}), (\ref{c42})and
(\ref{c42b})  leed to the expression
\begin{equation}
\label{c43}
 \ln{C_{LMC}} = S + \ln{D} = \ln{\left (\frac{3}{4}\right )} - 1
- \int_{-1}^{1} \ln{\left (\cos{\left (\frac{n\pi u}{2}\right )}du
\right )}
\end{equation}
and
\begin{equation}
\label{c43b}
 \ln{C_{LMC}} = S + \ln{D} = \ln{\left (\frac{3}{4}\right )} - 1
- \int_{-1}^{1} \ln{\left (\sin{\left (\frac{n\pi u}{2}\right )}du
\right )}
\end{equation}
for the even and odd cases, respectively. So the  complexity for a
particle in a one-dimentional box contains, except the quantum
number $n$, no parameter not only in the ground- but excited
states, as well. It is possible to extend this result for the
particle in a spherical box, $PIASB$ represented by the radial
Schr\"odinger equation
\begin{equation}
\label{radScheq} \frac{d^2R_{nl}}{dr^2}+\frac
2r\frac{dR_{nl}}{dr}+\left[2E-\frac{l(l+1)}{r^2}\right]R_{nl}=0
\end{equation}
where the radial wave function $R_{nl}(r)$ is given by
\begin{equation}
\label{radScheq1} R_{nl}=N j_{l}(r\sqrt{2E})
\end{equation}
where $j_{l}$ is the Bessel function of the first kind of order
${l}$ and N denotes the normalization constant.With  the Dirichlet
boundary condition imposed according to $R_{nl}(r_c)=0$, one
obtains through the condition $j_{l}(r_c\sqrt{2E})=0$ or
$(r_c\sqrt{2E})= u_{l,k}$, the energy levels given by
$E_{k,l}=\frac{{u_{l,k}}^2}{2{r_{c}}^2}$, where $u_{k,l}$ denotes
the $k^(th)$ zero of $j_{l}$. For the ground state, one obtains
for the disequilibrium
\begin{equation}
\label{cc17} D =
\frac{Si({2\pi})-\frac{Si({4\pi})}{2}}{(r_c)^{3}}=
\frac{0.6720709}{(r_c)^{3}} .
\end{equation}
The Shannon entropy can be written as
\begin{equation}
\label{cc20} S_r = \ln({8\pi}{r_c}^3)-3 +\frac{Si({2\pi})}{\pi}
\end{equation}
From Eqs. (\ref{cc17}) and (\ref{cc20}) we obtain the relation
\begin{equation}
\label{cc18} C_{LMC} = 1.3207394 ,
\end{equation}
which is constant with respect to the radius of confinement
${r_c}$. A similar analysis in the momentum space using the
Fourier transform of the position space wave function in the case
of PIASB leads to $C_{LMC}$ = 1.517215.
 As our second example, we consider free linear harmonic oscillator
with potential $V=1/2k x^2$. Then the ground-state density has the
form
\begin{equation}
\label{c15} \varrho =  \frac{k^{1/4}}{\pi^{1/2}} \exp{(-k^{1/2}
x^2)}.
\end{equation}
We can immediately obtain the Shannon entropy:
\begin{equation}
\label{c16} S =  -\ln{\left (\frac{k^{1/4}}{\pi^{1/2}}\right)} +
\frac12
\end{equation}
and the disequilibrium:
\begin{equation}
\label{c17} D =  \frac{k^{1/4}}{(2\pi)^{1/2}}  .
\end{equation}
From Eqs. (\ref{c16}) and (\ref{c17}) we are led to the relation
\begin{equation}
\label{c18}
 \ln{C_{LMC}} = S + \ln{D} =  \frac12(1- \ln{2})  .
\end{equation}
Thus the  complexity of the linear harmonic oscillator in the
ground state is
\begin{eqnarray}
\label{c18b}
 C_{LMC} = e^{\frac12 (1-\ln{2})} .
\end{eqnarray}
This is a remarkable result as it is a constant, independent of k.

The third example is the free hydrogen atom, or hydrogen-like
atomic ions.
 The ground-state density takes the form
\begin{equation}
\label{c19} \varrho =  \frac{Z^3}{\pi} \exp{(-2Zr)}  ,
\end{equation}
where $Z$ is the atomic number. The Shannon entropy can be written
as
\begin{equation}
\label{c20} S = 3 -\ln{\left (\frac{Z^3}{\pi}\right )} ,
\end{equation}
while the disequilibrium takes the form:
\begin{equation}
\label{c21} D =  \frac{Z^3}{8\pi}  .
\end{equation}
Eqs. (\ref{c20}) and (\ref{c21})  lead to the expression
\begin{equation}
\label{c22}
 \ln{C_{LMC}} = S + \ln{D} = 3 (1- \ln{2})  ,
\end{equation}
so the  complexity of hydrogen-like atomic ions in the ground
state is
\begin{eqnarray}
\label{c22b}
 C_{LMC} = e^{3 (1-\ln{2})} .
\end{eqnarray}
Here, again there is no parameter in the result.

In the Moshinsky model
 two electrons with antiparallel spins
 interact  harmonically in isotropic harmonic confinement.
 The Hamiltonian has the form
\begin{eqnarray}
\label{harm1} H = \frac12 \left(-\nabla^2_1 + r_1^2 \right) +
\frac12 \left(-\nabla^2_2 + r_2^2 \right) + \frac12 K r^2_{12}  ,
\end{eqnarray}
where
\begin{eqnarray}
\label{harm2} {\bf r}_{12} = {\bf r}_1 -{\bf r}_2
\end{eqnarray}
and $K$ is the coupling constant. Introducing relative
(Eq.(\ref{harm2})) and centre of mass coordinates
\begin{eqnarray}
\label{harm3} {\bf R} = \frac12 \left({\bf r}_1 +{\bf r}_2\right)
\end{eqnarray}
the Schr\"odinger equation can be separated leading to the
equations
\begin{eqnarray}
\label{harm4} \left(-\frac12 \nabla^2_R +2 R^2 \right) \Psi^{CM} =
E^{CM} \Psi^{CM}
\end{eqnarray}
and
\begin{eqnarray}
\label{harm5} \frac12 \left(-\nabla^2_{r_{12}} +\frac12 (\frac12 +
K) r^2_{12}
 \right) \Psi^{RM}
= E^{RM} \Psi^{RM} .
\end{eqnarray}
The ground-state density of the Moshinsky atom has the form
\cite{mhnr}
\begin{eqnarray}
\label{harm22} \varrho({\bf r}) = a \exp{(- b r^2)}  ,
\end{eqnarray}
where
\begin{eqnarray}
\label{harm7} a = 2 \left(\frac{b}{\pi}\right)^{3/2}  ,
\end{eqnarray}
\begin{eqnarray}
\label{harm9} b = \frac{2 \alpha - 1 }{ \alpha}
\end{eqnarray}
and
\begin{eqnarray}
\label{harm8} \alpha = \frac12 \left( 1 + (1+2K)^{1/2} \right).
\end{eqnarray}

From  Eqs. (\ref{c7}), (\ref{c9}), (\ref{c10}) and (\ref{harm22})
we arrive at the expression
\begin{eqnarray}
\label{r28}
 S +  \ln{D} =\frac32 (1-\ln{2})   .
\end{eqnarray}
Consequently, the  complexity of the Moshinsky atom in the ground
state is
\begin{eqnarray}
\label{r30b}
 C_{LMC} = e^{\frac32 (1-\ln{2})} .
\end{eqnarray}
This is again remarkable as it is a constant, independent of the
coupling constant. In all the analytical results treated in this
section, the parameter-free result for $C_{LMC}$ is is consequence
of the homogeneous property of the corresponding potential. This
has been proven earlier in the cases of the composite information
measures \cite{katriel} and for complexity \cite{Sen2}. When the
homogeneous property of the potential is lost, for example, under
the confined conditions as we shall see in the next section, the
$C_{LMC}$ values become
dependent on some parameter of the potential.\\
\section{ Numerical results on confined model systems}
The spherically confined Coulomb potential was originally
introduced \cite{Michels1} in order to study the effects of high
pressure on the dipole polarizability of hydrogen atom.In this
model, the confining potential is defined in terms of an
impenetrable spherical cavity of a certain radius of confinement,
$R$. The spherically confined Coulomb potential for hydrogen-like
atoms, for example, is given by
\begin{eqnarray}\label{eq:eq3}
V(r)&=&-\frac{Z}{r} \quad {\rm for}~~r< R,\nonumber \\
&=&{\infty}\quad {\rm for}~~ r\geq R.
\end{eqnarray}
We have carried out the calculations of $C_{LMC}$ corresponding to
the following spherically confined model systems (a) particle in a
spherical box (PIASB) (b)spherically confined isotropic harmonic
oscillator (SCIHO),(c)spherically confined hydrogen atom (SCHA),
all in position and momentum space and the two electron systems of
(d) a pair of electrons and (e) He atom in position space. For the
two electron systems we have employed the accurate 23-term and
70-term Hylleraas wave functions for the (d) and (e) cases,
respectively. The numerical procedure adopted by us has been
described in the literature \cite{Sen1,Sen2,Sen3} and will not be
repeated here. Our results on the calculations of $C_{LMC}$ for
systems (a)-(d) as a function of the radius of confinement, $R$,
are displayed in Fig. 1.

For the PIASB cases (position and momentum space) , the
homogeneous character of the potential leads to a constant value
of $C_{LMC}$ \cite{Sen2,katriel}. In all other cases, the absence
of homogeneous property of the potential gives rise to a nonlinear
dependence of $C_{LMC}$ with $R$. The range of confinements
considered by us presents a good test of the lower bound on
$C_{LMC}$ obtained in Eq.(9). As is evident from Fig. 1, this
bound is obeyed for all $R$ values across a diverse set of quantum
confined systems. The \emph{general} lower bound on $C_{LMC}$
derived in the present work is a significant addition to the
characteristics of this statistical measure.
\section{Conclusions}
 In summary, we derived a general lower bound for the shape complexity
measure of Lopez et al. \cite{lopez1995,lopez2002} We also
calculated the LMC measure
 for simple model systems:  the harmonic oscillator, the
hydrogen atom and  two-electron 'entangled artificial' atom
proposed by Moshinsky. We obtained the remarkable result that
 the LMC  measure is constant, that is, it does not depend on any parameter.
 This is characteristic of the homogeneous property of the
 corresponding potentials. A representative set of confined Coulomb potentials
 have been used to numerically test the general validity of the lower bound
 obtained on $C_{LMC}$. Due to the non-homogeneous property of the
 potentials in these cases, $C_{LMC}$ becomes dependent on the
 parameters of the potential.
\section*{Acknowledgements}

This work was  written in the frame of the Bilateral
Intergovermental Scientific and Technological Cooperation between
Hungary and India sponsored by the Research an Technological
Innovation Foundation and the Ministry of Science and Technology.
\'AN acknowledges grant OTKA No. T\,67923. KDS is grateful to
Jorge Garza and Rubicila Vargas for kind hospitality during the
course of confined systems meeting held at UAM-I Mexico, in April
2009. We dedicate this work to the fond memory of Professor Marcos
Moshinsky, UNAM, Mexico.

\begin{figure}
\includegraphics[scale=0.75]{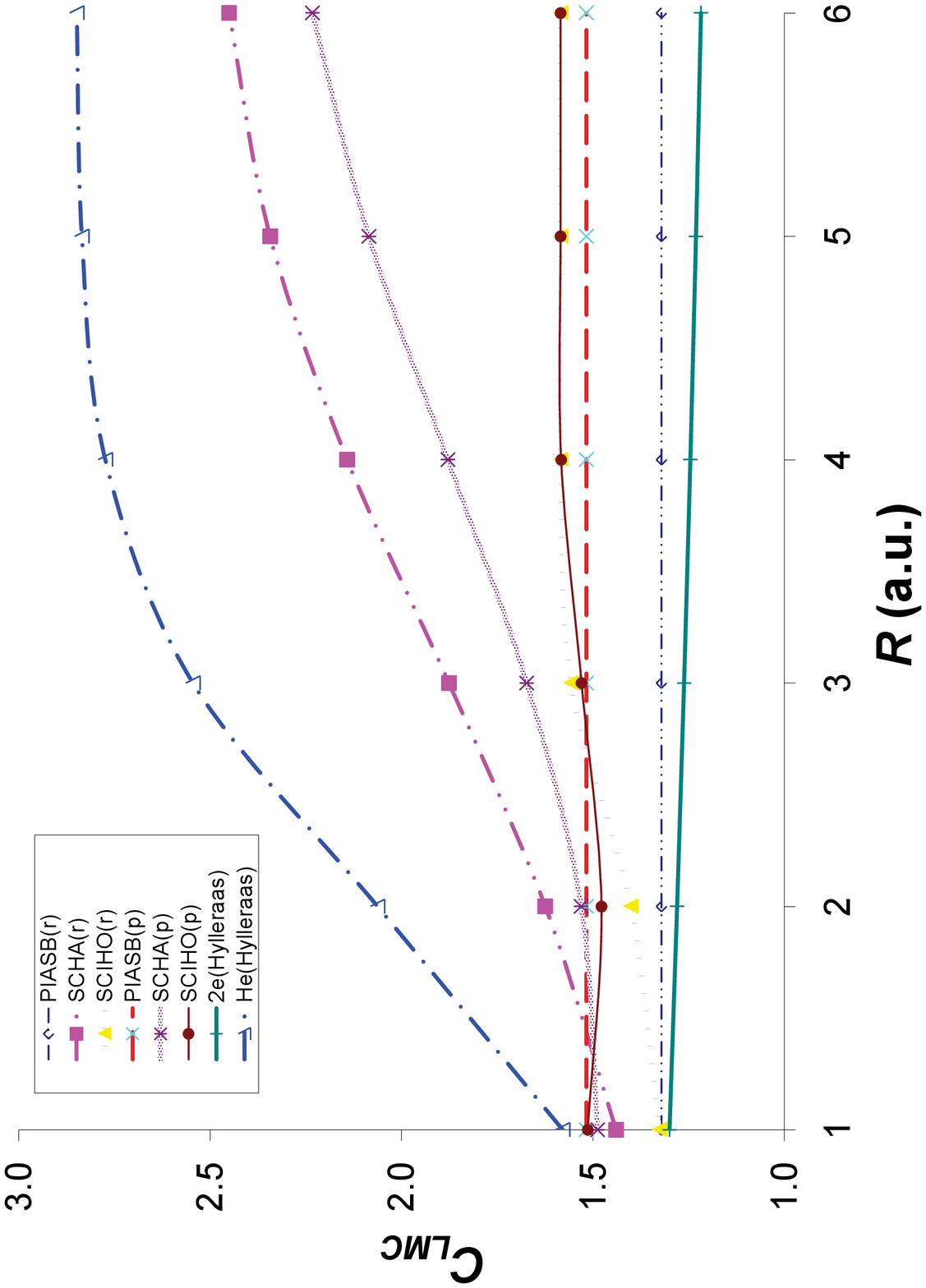}
\caption{Variation of $C_{LMC}$ corresponding to the spherically
confined model systems in position and momentum space for (a)
particle in a spherical box (PIAB) (b) isotropic harmonic
oscillator (SCIHO),(c) hydrogen atom (SCHA), and in position space
for (d) a pair of electrons and (e) He atom. For (d) 23-term , and
for (e) 70-term, Hylleraas wave functions have been employed.}

\end{figure}

\end{document}